\documentclass[prb,amsmath,amssymb,superscriptaddress,twocolumn]{revtex4}
\usepackage{float}
\usepackage{amsmath}

\usepackage{amssymb}
\usepackage{amsfonts}
\usepackage{euscript}
\usepackage{enumerate}
\usepackage{hhline}
\usepackage{pslatex}
\usepackage{tabularx}
\usepackage[usenames,dvipsnames]{xcolor}

\usepackage{graphicx}

\usepackage{dcolumn}
\usepackage{bm}
\usepackage[sort&compress]{natbib}

\makeatletter
\renewcommand{\@biblabel}[1]{#1. }
\renewcommand{\@dotsep}{500}
\renewcommand{\@pnumwidth}{0em}
\renewcommand{\l@figure}[2]{
\@dottedtocline{1}{1.5em}{2em}{Figure #1}{}\vspace{15pt}}

\newcommand{\ks}[1]{\textcolor{blue}{#1}}

\usepackage[normalem]{ulem}

\begin{document}

\title{Milliwatt-threshold visible-telecom optical parametric oscillation using silicon nanophotonics}

\author{Xiyuan Lu}\email{xiyuan.lu@nist.gov}
\affiliation{Microsystems and Nanotechnology Division, Physical Measurement Laboratory, National Institute of Standards and Technology, Gaithersburg, MD 20899, USA}
\affiliation{Maryland NanoCenter, University of Maryland,
College Park, MD 20742, USA}
\author{Gregory Moille}
\affiliation{Microsystems and Nanotechnology Division, Physical Measurement Laboratory, National Institute of Standards and Technology, Gaithersburg, MD 20899, USA}
\affiliation{Maryland NanoCenter, University of Maryland,
College Park, MD 20742, USA}
\author{Anshuman Singh}
\affiliation{Microsystems and Nanotechnology Division, Physical Measurement Laboratory, National Institute of Standards and Technology, Gaithersburg, MD 20899, USA}
\affiliation{Maryland NanoCenter, University of Maryland,
College Park, MD 20742, USA}
\author{Qing Li}
\affiliation{Microsystems and Nanotechnology Division, Physical Measurement Laboratory, National Institute of Standards and Technology, Gaithersburg, MD 20899, USA}
\affiliation{Maryland NanoCenter, University of Maryland,
College Park, MD 20742, USA}
\affiliation{Electrical and Computer Engineering, Carnegie Mellon University, Pittsburgh, PA 15213, USA}
\author{Daron A. Westly}
\affiliation{Microsystems and Nanotechnology Division, Physical Measurement Laboratory, National Institute of Standards and Technology, Gaithersburg, MD 20899, USA}
\author{Ashutosh Rao}
\affiliation{Microsystems and Nanotechnology Division, Physical Measurement Laboratory, National Institute of Standards and Technology, Gaithersburg, MD 20899, USA}
\affiliation{Maryland NanoCenter, University of Maryland,
College Park, MD 20742, USA}
\author{Su-Peng Yu}
\affiliation{Time and Frequency Division, Physical Measurement Laboratory, National Institute of Standards and Technology, Boulder, CO 80305, USA}
\affiliation{Department of Physics, University of Colorado, Boulder, CO 80309, USA}
\author{Travis C. Briles}
\affiliation{Time and Frequency Division, Physical Measurement Laboratory, National Institute of Standards and Technology, Boulder, CO 80305, USA}
\affiliation{Department of Physics, University of Colorado, Boulder, CO 80309, USA}
\author{Tara Drake}
\affiliation{Time and Frequency Division, Physical Measurement Laboratory, National Institute of Standards and Technology, Boulder, CO 80305, USA}
\affiliation{Department of Physics, University of Colorado, Boulder, CO 80309, USA}
\author{Scott B. Papp}
\affiliation{Time and Frequency Division, Physical Measurement Laboratory, National Institute of Standards and Technology, Boulder, CO 80305, USA}
\affiliation{Department of Physics, University of Colorado, Boulder, CO 80309, USA}
\author{Kartik Srinivasan} \email{kartik.srinivasan@nist.gov}
\affiliation{Microsystems and Nanotechnology Division, Physical Measurement Laboratory, National Institute of Standards and Technology, Gaithersburg, MD 20899, USA}
\affiliation{Joint Quantum Institute, NIST/University of Maryland, College Park, MD 20742, USA}
\date{\today}

\begin{abstract}
     \noindent \textbf{The on-chip creation of coherent light at visible wavelengths is crucial to field-level deployment of spectroscopy and metrology systems. Although on-chip lasers have been implemented in specific cases, a general solution that is not restricted by limitations of specific gain media has not been reported. Here, we propose creating visible light from an infrared pump by widely-separated optical parametric oscillation (OPO) using silicon nanophotonics. The OPO creates signal and idler light in the 700~nm and 1300~nm bands, respectively, with a 900~nm pump. It operates at a threshold power of (0.9 $\pm$ 0.1) mW, over 50$\times$ smaller than other widely-separated microcavity OPO works, which have only been reported in the infrared. This low threshold enables direct pumping without need of an intermediate optical amplifier. We further show how the device design can be modified to generate 780 nm and 1500 nm light with a similar power efficiency. Our nanophotonic OPO shows distinct advantages in power efficiency, operation stability, and device scalability, and is a major advance towards flexible on-chip generation of coherent visible light.}
\end{abstract}

\maketitle
On-chip generation of coherent light at visible frequencies is critical for miniaturization and field-level deployment for spectroscopy and metrology, for example, wavelength-stabilized reference lasers based on atomic vapors \cite{Hummon2018} and optical atomic clocks \cite{Ludlow2015}. One approach is to develop on-chip lasers directly using III-V semiconductors~\cite{Sun2016}, but the wavelength coverage is limited by the available gain media and requires nontrivial heterogeneous integration to be compatible with a silicon chip. Another approach is to use nonlinear optics to create light at new frequencies from existing lasers. There are many second/third-order ($\chi^{(2)}$/$\chi^{(3)}$) nonlinear optical processes for this purpose, including optical parametric oscillation (OPO), second/third harmonic generation (SHG/THG), sum frequency generation (SFG), and stimulated four-wave mixing/optical parameteric amplification (StFWM/OPA) \cite{Boyd2008,Agrawal2007}. Among these processes, OPO is uniquely suitable to generate coherent light over a wide spectral range, because the generated light is not limited to harmonics of pump frequencies (unlike SHG/THG), and only one pump laser is required (unlike SFG/StFWM/OPA). Historically, $\chi^{(2)}$ OPO has been particularly efficient in creating coherent light across wide spectral ranges. Half a century ago, coherent OPO light from 970 nm to 1150 nm was generated by a 529 nm pump laser in a LiNbO$_3$ crystal \cite{Giordmaine1965}. Later, the OPO signal wavelength was brought into the visible (from 537 nm to 720 nm) by a 308 nm pump laser in a BaB$_2$O$_4$ crystal~\cite{Ebrahimzadeh1989}. Nowadays, $\chi^{(2)}$ OPO systems have become a laboratory workhorse tool in the generation of coherent, tunable visible light, particularly when pumped by a Ti:Sapphire laser \cite{Moulton1986}.  However, nanophotonic implementations of OPO that can reach visible wavelengths are still lacking.

\begin{figure*}[htbp]
\centering\includegraphics[width=0.9\linewidth]{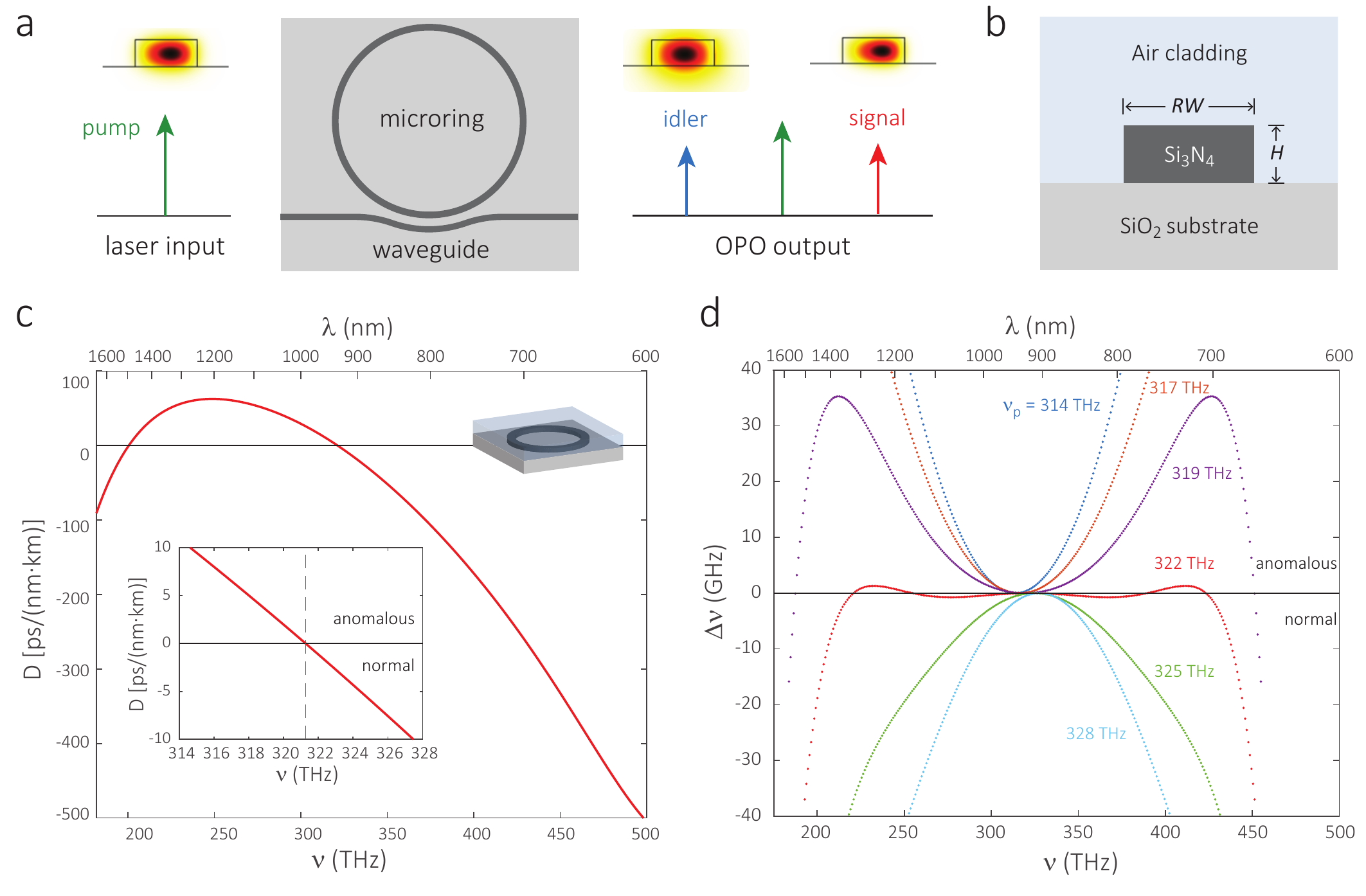}
\caption{{\bf Design of a nanophotonic visible-telecom optical parametric oscillator.} \textbf{a}, Schematic indicating that the microring device uses cavity-enhanced degenerate four-wave mixing (dFWM) to generate signal and idler light that have frequencies widely separated from the input pump. All interacting modes (pump, signal, and idler) are fundamental transverse-electric modes (TE1), with their dominant electric field components shown in insets. The input pump and the output signal and idler are all coupled with the same waveguide in this scheme. \textbf{b}, A cross-section view of the microring shows the air cladding and silicon dioxide substrate, and two key geometric parameters, ring width ($RW$) and height ($H$). These two parameters, together with the ring outer radius ($RR$), unambiguously determine the microring dispersion. \textbf{c}, Dispersion curve ($D$) of a typical geometry, with $RR$ = 23 $\mu$m, $RW$ = 1160 nm, and $H$ = 510 nm. $D=0$ when the pump frequency $\nu_{\text{p}}$ is $\approx$ 321.7 THz (932.5 nm), as shown in the zoomed-in inset. The dispersion is anomalous ($D>0$) when $\nu_{\text{p}}$ is smaller, and normal ($D<0$) when $\nu_{\text{p}}$ is larger. \textbf{d}, Frequency mismatch ($\Delta \nu$) for dFWM for the geometry in (c) at various values of $\nu_{\text{p}}$. When the pump is slightly normal at 322 THz (red), there are two cases in which signal and idler modes are phase-/frequency-matched, with both suitable for widely-separated OPO. $\Delta \nu$ is calculated for specific mode number ($m$) sets, because dFWM requires the phase-matching condition to be satisfied, i.e., $m_{\rm s} + m_{\rm i} = 2 m_{\rm p}$. The mode frequency for each mode number is calculated for the geometry in (c) by the finite-element method.}
\label{Fig1}
\end{figure*}

In this work, we report a nanophotonic $\chi^{(3)}$ OPO for on-chip visible light generation. We use $\chi^{(3)}$ rather than $\chi^{(2)}$ for two important reasons. First, $\chi^{(3)}$ processes, unlike $\chi^{(2)}$ processes, are naturally compatible with silicon photonics. Second, as the $\chi^{(3)}$ OPO consumes two pump photons for each generated signal/idler pair, visible wavelengths can be reached through an easily available infrared pump laser, in contrast to a $\chi^{(2)}$ OPO which needs a UV pump laser. Although ideas for how to achieve widely-separated $\chi^{(3)}$ OPO have been demonstrated in photonic crystal fiber based systems~\cite{Chen2005,Deng2005} and then introduced to silicon nanophotonics theoretically over a decade ago~\cite{Lin2008}, because of the challenging dispersion engineering requirements, such nanophotonic OPO has not been experimentally demonstrated. This is in contrast to other wide-band nonlinear processes, e.g., nanophotonic SHG/THG, which have been extensively reported \cite{Levy2011, Guo2016a, Lin2016, Fujii2017, Vahala2007, Surya2018, Chang2019}. Recently, widely separated OPO has been achieved in whispering-gallery mode (WGM) platforms with larger footprints, including crystalline ${\rm MgF_2}$ microcavities \cite{Sayson2017,Fujii2019,Sayson2019} and ${\rm SiO_2}$ microtoroids \cite{Fujii2017}, but the threshold powers are relatively large and the spectra of the OPO output have been restricted to the infrared.

To demonstrate nanophotonic visible-telecom OPO, we use the silicon nitride (Si$_3$N$_4$) platform, whose advantageous characteristics for  silicon-based nonlinear nanophotonics \cite{Moss2013}, including octave-spanning frequency combs \cite{Okawachi2011, Li2017, Karpov2018}, frequency conversion/spectral translation \cite{Li2016, Singh2019, Lu2019B}, entanglement generation \cite{Lu2019A}, and clustered frequency comb generation \cite{Matsko2016, Huang2017}, has by now been well-established. Here we show, for the first time, on-chip OPO with signal and idler at visible and telecom frequencies, for example, 419.8 THz (714.6 nm) and 227.8 THz (1316.9 nm), respectively. The OPO process is power efficient due to nanophotonic confinement and strong spatial mode overlap, and has an ultra-low threshold power of (0.9 $\pm$ 0.1) mW. In contrast to recent microresonator OPO work that use between 50 mW and 380 mW of pump power to achieve widely-separated signal and idler in the infrared \cite{Fujii2017,Sayson2017,Fujii2019,Sayson2019}, our devices use only milliwatt-level power, without intermediate optical amplifiers, to achieve widely-separated signal and ilder in the visible and telecom, respectively. We further show that the OPO frequencies can be readily controlled by changing the device geometry. In particular, we demonstrate OPO with signal and idler at 383.9 THz (781.4 nm) and 202.1 THz (1484 nm) by pumping at 293.0 THz (1024 nm). This signal wavelength is suitable for Rubidium vapor, and the pump wavelength is accessible from compact semiconductor chip lasers.

\noindent \textbf{Design principles}
Our OPO devices are based on cavity-enhanced degenerate four-wave mixing (dFWM), which requires conservation of both momentum and energy for the interacting optical modes \cite{Vahala2004}. For the same mode family, momentum conservation is simplified to conservation of the azimuthal mode number, that is, $\Delta m = m_\text{s}+m_\text{i}-2m_\text{p} = 0$, where the subscripts $s,i,p$ denote signal, idler, and pump, respectively. Energy conservation requires the central frequencies of the cavity modes to have a mismatch ($\Delta \nu = \nu_\text{s}+\nu_\text{i}-2\nu_\text{p}$) within the cavity linewidths, that is, $|\Delta \nu| < \nu_k/Q_k$, where $k=s,i,p$ and $Q_k$ is the loaded quality factor for the $k$ mode.
We note that achieving such phase and frequency matching across visible and telecom bands has only been demonstrated recently in silicon nanophotonics for photon-pair generation \cite{Lu2019A} and spectral translation \cite{Lu2019B}, where a mode splitting approach \cite{Lu2014} enables the identification of specific azimuthal modes separated by hundreds of THz. We employ a similar approach here, focusing on fundamental transverse electric (TE1) modes only, which have high-$Q$, strong modal confinement ($\bar{V}$), and good mode overlap ($\eta$).  These attributes are essential for achieving low-threshold operation, as discussed in the Supplementary Information Section I.

However, the above design principles do not guarantee that the targeted wide-band OPO process will occur. Critically, the targeted process also has to win over all other competing processes that are matched in phase and frequency, including OPO in the pump band~\cite{Lu2019A, Lu2019B}, clustered frequency combs in the signal and idler bands \cite{Fujii2017, Fujii2019, Matsko2016, Huang2017}, and other nonlinear processes (e.g., stimulated Raman scattering \cite{Sayson2017} and third-harmonic generation \cite{Fujii2017}). For example, recent work reporting telecom-to-visible spectral translation via stimulated dFWM did not exhibit widely-separated OPO, because without the seed telecom light, close-to-pump OPO processes dominate~\cite{Lu2019B}. Thus, unlike previous work in wide-band silicon nonlinear nanophotonics~\cite{Okawachi2011,Li2017,Karpov2018,Li2016,Lu2019A,Lu2019B}, visible-telecom OPO faces a more stringent requirement not only on enhancing the process of interest, but also on suppressing all competing processes at the same time.

In particular, OPO in the pump band can be suppressed if the pump modes are in the normal dispersion regime~\cite{Lin2008}, which corresponds to a negative dispersion parameter ($D$). $D = - \frac{\lambda}{c} \frac{d^2 \bar{n}}{d\lambda ^2}$, where $c$, $\lambda$, and $\bar{n}$ represent the speed of light, vacuum wavelength, and effective mode index, respectively~\cite{Agrawal2007}. $D < 0$ is equivalent to $\Delta \nu < 0$ when the signal and idler modes are near the pump mode. Therefore, we need to design the device to have $\Delta \nu < 0$ when signal and idler are near the pump, and $\Delta \nu = 0$ when signal and idler are widely separated.

\noindent \textbf{Numerical simulations} We use the aforementioned design principles to guide numerical simulations for the widely-separated OPO. Figure~\ref{Fig1}(b) shows a cross-sectional view of the microring. The Si$_3$N$_4$ core has a rectangular cross-section, described by ring width ($RW$), thickness ($H$), and ring radius ($RR$). We use these parameters to tailor the geometric contribution to the dispersion.  We note that self-/cross-phase modulation is negligible in our device, so that we can use the natural cavity frequencies to design our OPO (See Supplementary Information Section I).

Figure~\ref{Fig1}(c) shows the dispersion parameter of a device with $RW$ = 1160 nm, $H$ = 510 nm, and $RR$ = 23 $\mu$m, where the zero dispersion frequency (ZDF) is at $\approx$ 321 THz. The dispersion is anomalous for smaller frequencies and normal for larger frequencies. The frequency mismatch ($\Delta \nu$) is plotted (Fig.~\ref{Fig1}(d)) with pump frequency ($\nu_{\text{p}}$) ranging from 314 THz to 328 THz. When $\nu_{\text{p}}=322$~THz, nearby modes show an overall small normal dispersion, and there are two widely-separated mode pairs that are frequency-matched ($\Delta \nu = 0$). In contrast, larger values of $\nu_{\text{p}}$ have large normal dispersion and do not lead to widely-separated OPO. Smaller $\nu_{\text{p}}$ may allow widely-separated OPO (e.g., 319 THz case), but the anomalous dispersion around the pump results in several close-band competitive OPO processes, making widely-separated OPO unavailable in general.

\begin{figure*}[t]
\centering\includegraphics[width=0.9\linewidth]{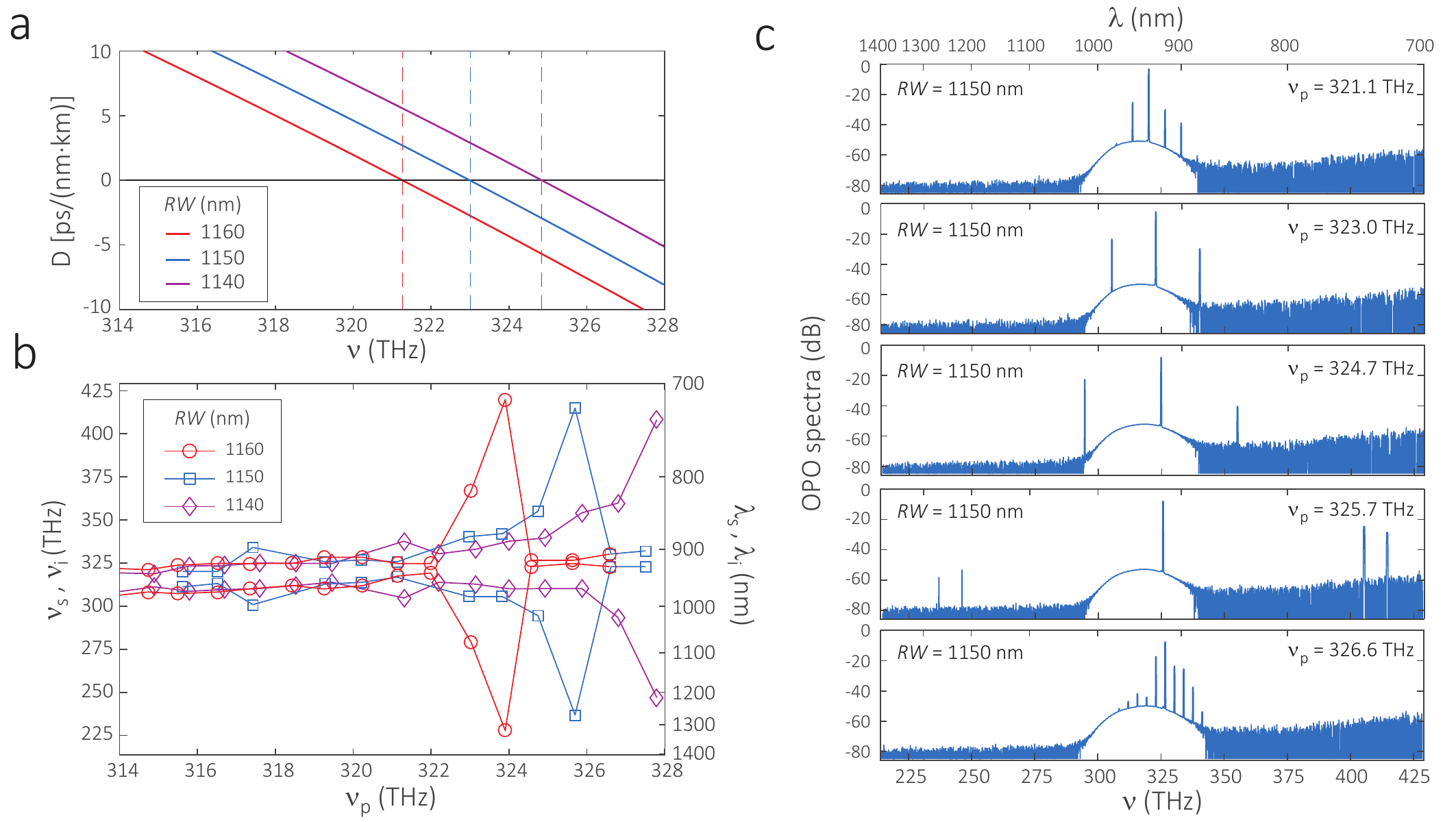}
\caption{{\bf OPO frequencies critically depend on ring width ($RW$) and pumping frequency ($\nu_{\rm p}$)}. \textbf{a}, Simulated dispersion ($D$) curves for different $RW$, with other parameters specified in the caption of Fig.~\ref{Fig1}. The zero dispersion frequency (ZDF) blueshifts with decreasing radius. \textbf{b}, Experimentally recorded OPO output (signal and idler) frequencies (left axis, $\nu_\text{s}$ and $\nu_\text{i}$) and wavelengths (right axis, $\lambda_\text{s}$ and $\lambda_\text{i}$) of the aforementioned geometries when $\nu_\text{p}$ is varied around the ZDF. Widely-separated OPO occurs when the dispersion is slightly normal, as suggested by Fig.~\ref{Fig1}, because potential close-band OPO processes are inhibited. \textbf{c}, OPO spectra for the $RW$ = 1150 nm device when $\nu_\text{p}$ is varied. When scanning $\nu_\text{p}$ from a mode in the anomalous region to one in the normal region, the spectral separation of the OPO signal and idler increases from 9 THz to 37 THz, 61 THz, and 178 THz, and finally decreases to 7 THz (from top to bottom). On the y axis, 0 dB is referenced to 1 mW, i.e., dBm.}
\label{Fig2}
\end{figure*}

We also simulate devices that have different $RW$ but the same $RR$ and $H$, with the dispersion plotted in Fig.~\ref{Fig2}(a). When $RW$ increases from 1140 nm to 1160 nm, the ZDF redshifts from 325 THz to 321 THz, remaining within our laser scanning range. We thus have a prescription for geometries to experimentally observe the transition from close-band to widely-separated OPO processes. For details regarding the parametric sensitivity in dispersion engineering, please refer to Supplementary Information Section III.

\noindent \textbf{Experimental measurements}
We fabricate devices (see Supplementary Information Section IV) with fixed $H$ and varying $RW$, and characterize them as a function of $\nu_{\text{p}}$ near the ZDF. The results are summarized in Fig.~\ref{Fig2}(b)-(c). The output OPO spectra are recorded by an optical spectrum analyzer (OSA), while the pump is scanned for modes that transit from anomalous to normal dispersion, with an example in Fig.~\ref{Fig2}(c) for fixed $RW$=1150~nm. The OPO signal and idler frequencies for all $RW$ and $\nu_{\text{p}}$ are then plotted in Fig.~\ref{Fig2}(b). $\nu_{\text{p}}$ for OPO with the widest separation redshifts from 327.8 THz ($RW=1160$~nm) to 325.7 THz ($RW=1150$~nm) and 323.8 THz ($RW=1140$~nm), following the shift in device dispersion. Focusing again on the $RW$ = 1150 nm spectra for several different $\nu_{\text{p}}$ (Fig.~\ref{Fig2}(c)), we clearly observe the trend predicted previously when tuning $\nu_{\text{p}}$ from anomalous to normal. When the pump dispersion is anomalous, OPO signal and idler bands are closely spaced around the pump (top panel in Fig.~\ref{Fig2}(c)). When the pump dispersion is slightly normal, the OPO signal and idler have increasingly large spectral separation as $\nu_{\text{p}}$ increases (2$^\text{nd}$ to 4$^\text{th}$ panel in Fig.~\ref{Fig2}(c)). However, when the pump dispersion is too normal, no widely-separated OPO is observed, and only very close-band OPO is seen (the bottom panel in Fig.~\ref{Fig2}(c)). The $RW$ = 1160 nm device (red in Fig.~\ref{Fig2}(b)) has a similar trend but fewer pumping modes in the transition to the slightly normal region. This trend agrees with the prediction from Fig.~\ref{Fig1}(d), although the experimental $\nu_{\text{p}}$ is 2 THz larger than predicted, which is likely due to uncertainties in device fabrication.

\begin{figure*}[htbp]
\centering\includegraphics[width=0.9\linewidth]{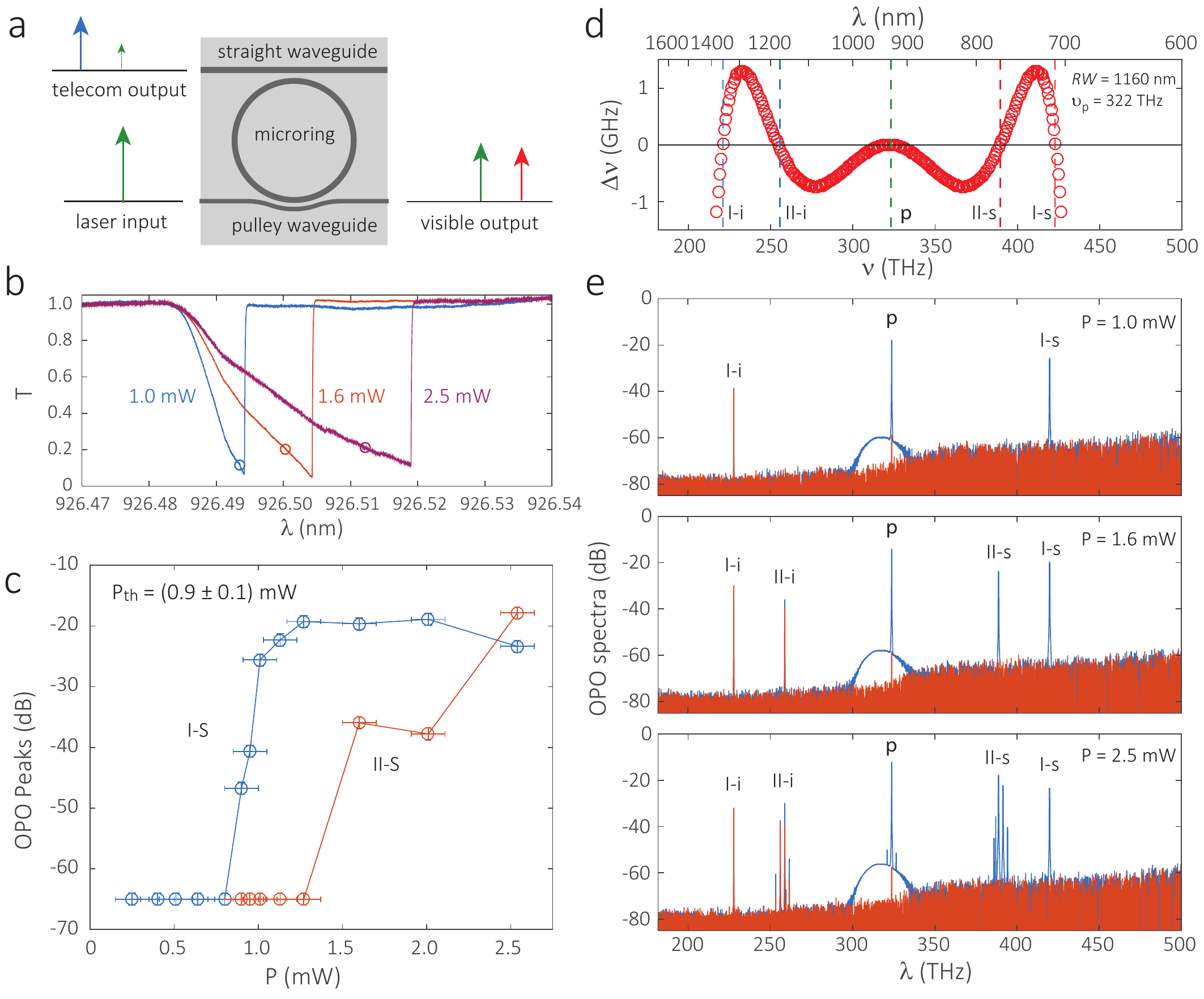}
\caption{{\bf Power dependence of the visible-telecom OPO.} \textbf{a}, When the OPO frequencies are separated widely into the visible-telecom regime, two waveguides are needed to couple the visible and telecom light efficiently. The straight waveguide (top) is used for out-coupling the telecom (idler) and the pulley waveguide (bottom) is for out-coupling the visible (signal). \textbf{b}, Transmission ($T$) traces for $\nu_{\text{p}}\approx$~322 THz show bistabilites with various pump powers ($P$). The open circles specify the laser detuning at various power for the OSA spectra in (e). \textbf{c}, OPO threshold power is only (0.9 $\pm$ 0.1) mW, measured by the power dependence of the OPO signal peak amplitudes. I-s and II-s denote signal peaks of two OPO tones at 419.8 THz and 388.7 THz in (e). Here, on the y axis, 0 dB is referenced to 1 mW, i.e., dBm.  Error bars are one standard deviation values due to fluctuations in optical path losses. \textbf{d}, Zoom-in frequency mismatch curve of Fig.~\ref{Fig1}(d) suggests two phase-/frequency-matched cases, where the signal/idler frequencies are labeled as I-s/I-i and II-s/II-i, respectively. The pump frequency is labeled as p (around 322 THz). \textbf{e}, OPO spectra as a function of pump power.  When the pump power is 1.0 mW, OPO I is above threshold and OPO II is below threshold, with I-s and I-i located around 419.8 THz (714.6 nm) and 227.8 THz (1316.9 nm), respectively, corresponding to a spectral separation of 192 THz. Next, when the pump power is 1.6 mW, both OPO I and II are above threshold and observable in the spectrum. II-s and II-i are located around 388.7 THz (771.8 nm) and 258.8 THz (1159.2 nm). The frequencies of both OPO I/II agree reasonably well with the theoretical prediction in (d). Finally, when the pump power is 2.5 mW, close-band FWM adjacent to OPO II is excited, because the modes adjacent to II have smaller frequency mismatch compared to those around I, as indicated by (d). On the y axis, 0 dB is referenced to 1 mW, i.e., dBm.}
\label{Fig3}
\end{figure*}

We focus on the $RW$ = 1160 nm device and study its power dependence at $\nu_\text{p}=322$~THz in Fig.~\ref{Fig3}. The OPO signal and idler have a spectral separation that is too large for a single waveguide (Fig.~\ref{Fig1}(a)) to out-couple both frequencies efficiently. We therefore use two waveguides to separate the coupling tasks, as shown in Fig.~\ref{Fig3}(a). The bottom pulley waveguide couples the pump and signal light together, while being cut-off at telecom wavelengths (Supplementary Information Section II). The top waveguide couples telecom light efficiently, but does not couple the pump and signal light due to a limited spatial overlap that prevents effective evanescent coupling. The combined coupling geometry is designed to have coupling $Q=(1-2)\times10^6$ for pump, signal, and idler modes. In experiment, we verify that the fabricated device has intrinsic $Q=(2-3)\times10^6$ and loaded $Q\approx1\times10^6$ for TE1 modes in the pump band, which corresponds to loaded cavity linewidths of $\approx$ 300 MHz. With such high $Q$, the device shows large thermal bistability at milliwatt pump powers, as shown in Fig.~\ref{Fig3}(b). For each pump power, we situate the pump detuning near the dip of the cavity resonance and measure the generated OPO spectrum. Three representative spectra are shown in Fig.~\ref{Fig3}(e), with pump detuning indicated by the open circles in Fig.~\ref{Fig3}(b). For 1~mW pump power at $\nu_\text{p}=323.8$~THz (926.5~nm), the top panel of Fig.~\ref{Fig3}(e) shows that a widely-separated OPO is generated with signal at 419.8 THz (714.6 nm) and idler at 227.8 THz (1317 nm). The signal-idler separation is 192~THz, comparable to the largest reported value for WGM resonators ($\approx$ 230 THz), where the idler frequency was inferred~\cite{Sayson2019} (signal and idler were both in the infrared). With an increased pump power of 1.6~mW, an additional pair is generated at 388.8 THz (771.6 nm) and 258.8 THz (1187 nm), as shown in the middle panel of Fig.~\ref{Fig3}(e). With a further increase in pump power to 2.5~mW, clustered combs are generated around the second signal-idler pair, while the first pair remains unaccompanied by other spectral tones (the bottom panel of Fig.~\ref{Fig3}(e)).

These two OPO pairs measured in experiment agree quite well with the theoretical predictions (Fig.~\ref{Fig3}(d)), where pair I is predicted to be at 423 THz (I-s) and 221 THz (I-i) and pair II at 389 THz (II-s) and 255 THz (II-i). The $m$ numbers of these modes are $\{$420, 383, 310, 237, 200$\}$ for $\{$I-s, II-s, p, II-i, I-i$\}$ (labeling scheme in  Fig.~\ref{Fig3}(d)). These mode numbers clearly satisfy phase-matching ($\Delta m = 0$). Moreover, the fact that the clustered comb is generated in the II pair, but not in the I pair, is not coincidental and can be explained as follows. All the mode pairs satisfying phase-matching are plotted in Fig.~\ref{Fig3}(d). Each mode is represented by an open circle and the cavity free spectral range (FSR) is $\approx$ 1 THz. Although both I and II satisfy frequency matching, the density of mode pairs (within a given range of frequency mismatch) around I and II are different. Because material dispersion is much larger at higher frequencies, the I pair exhibits larger dispersion and has sparser modes in the neighborhood of the tolerated frequency mismatch, which can be estimated by the cavity linewidth ($\approx$ 300 MHz). In other words, the modes near II are preferred for clustered comb generation considering both mode density and frequency matching. Moreover, because of the normal dispersion around the pump, there are no competitive processes in the pump band even at higher pump power (Fig.~\ref{Fig3}(e)). A power-dependence study (Fig.~\ref{Fig3}(c)) indicates a threshold of (0.9~$\pm$~0.1)~mW for the first set of OPO lines. The second OPO has a threshold of (1.5~$\pm$~0.2)~mW, while its subsequent clustered frequency comb has a threshold near 2.5~mW.

\noindent \textbf{OPO on a single widely-separated pair}
In the previous section, although close-band OPO with spectral tones near the pump are successfully suppressed, the generation of two OPO pairs with pair II eventually exhibiting a cluster of tones might be unwanted in applications. Here we show how the ring geometry can be tuned to achieve a dispersion that supports only one single set of widely-separated OPO tones.

\begin{figure*}[t!]
\centering\includegraphics[width=0.9\linewidth]{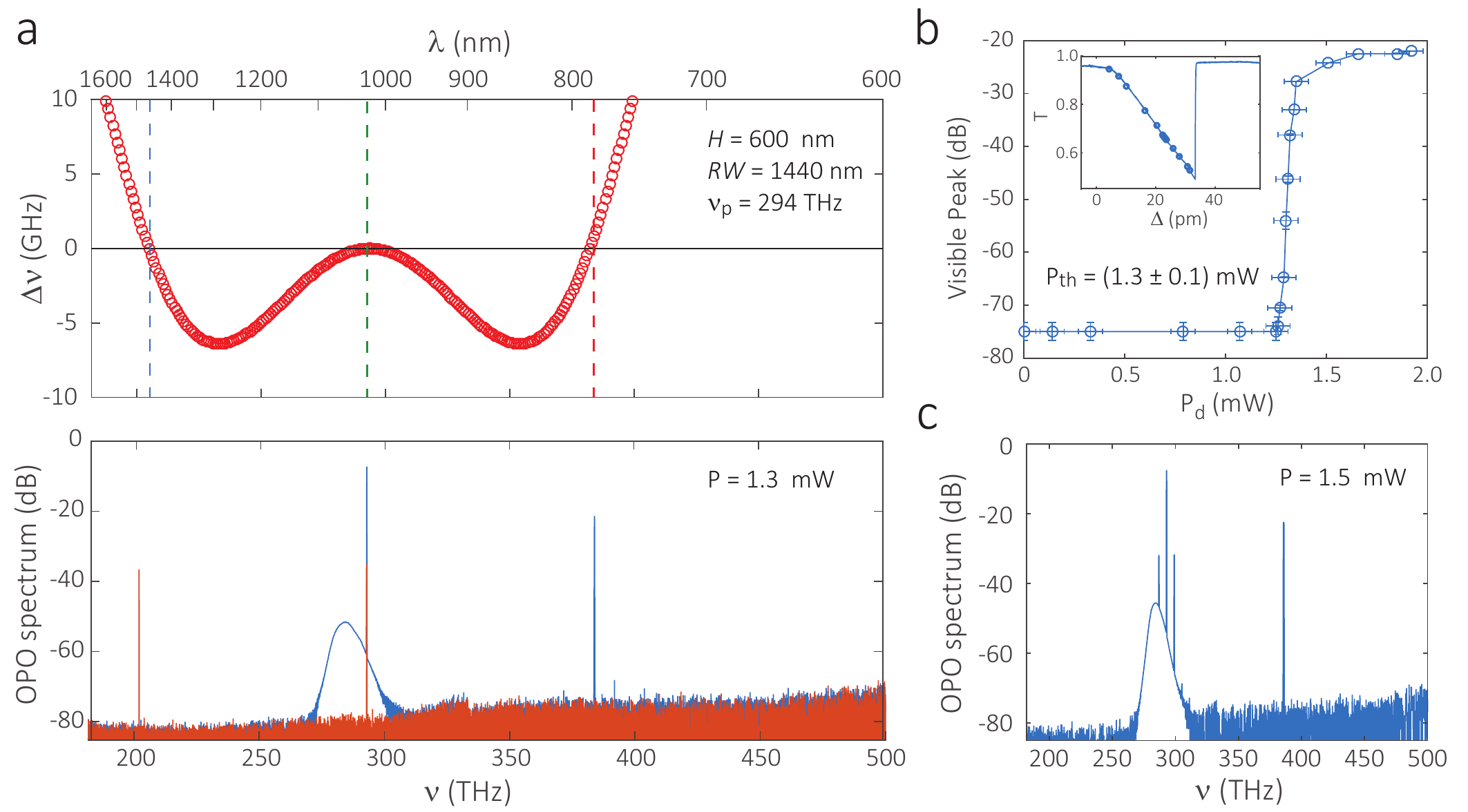}
\caption{{\bf OPO on a single widely-separated pair.} \textbf{a}, Top panel shows a microring design with only one phase-/frequency-matched widely-separated OPO pair. The microring has parameters of $H$ = 600 nm and $RW$ = 1440 nm. When the pump laser frequency $\nu_{\rm p}$ = 293.5 THz, the generated OPO is predicted to have only a single pair with frequencies of 205 THz and 382 THz. The bottom panel shows the experimental optical spectrum, which confirms that only a single widely-separated pair is generated at 202.1 THz (1484 nm) and 383.9 THz (781.4 nm) when $\nu_{\rm p}$ = 293.0 THz (1024 nm). Due to the large spectral separation, the device needs two waveguides to couple the OPO signal and idler, with spectra shown in red and blue, respectively. \textbf{b}, Threshold study of the OPO pair with various dropped pump power $P_\text{d} = P~(1-T)$, where transmission (T) is changed by the laser-cavity detuning ($\Delta$), as shown in the inset. The threshold power is (1.3 $\pm$ 0.1) mW. \textbf{c}, The dispersion is normal near the pump, as shown in (a), thereby disfavoring close-band OPO. However, when the pump power is sufficiently above threshold, the close-band OPO process begins to appears. This competitive OPO is much less efficient than the widely-separated OPO, but nevertheless needs further suppression for ideal operation. In the y axes of (b-c), 0 dB is referenced to 1 mW, i.e., dBm.}
\label{Fig4}
\end{figure*}

We calculate the device dispersion and OPO frequency mismatch for various ring widths using mode frequencies from FEM simulations. The top panel of Fig.~\ref{Fig4}(a) shows the key result where the device with $RW$ = 1440 nm is predicted to generate a visible-telecom OPO with signal and idler located at 384 THz (781 nm) and 204 THz (1470 nm), for a pump at 294 THz (1020 nm). This $H=600$~nm design has widely-separated frequency-/phase-matching mode pairs and normal dispersion near the pump, similar to the previous $H=510$~nm design (Fig.~\ref{Fig3}(d)). However, the $H=600$~nm design supports only one widely-separated OPO pair, and is also $\approx$ 5$\times$ more dispersive in the frequency bands of interest. These two properties together make this design better in suppressing competitive OPO processes. In the measured optical spectrum (the bottom panel of Fig.~\ref{Fig4}(a)), the fabricated device generates OPO with signal and idler at 383.9 THz (781.5 nm) and 202.1 THz (1484 nm), respectively, when pumping at 293.0 THz (1023.9 nm) with 1.3~mW pump drop power. The measured frequencies agree with theoretical prediction within 2 THz for all three modes. Moreover, only one pair of widely-separated tones is generated, as the simulation predicts. We note that the short wavelength OPO output is suited for spectroscopy of Rubidium vapor (1.5~nm wavelength tuning needed), and the telecom OPO output makes such a device potentially suitable for spectral translation \cite{Lu2019A}.

One unique property of our device is its operation stability, that is, OPO works at a continuous detuning of the pump. This stability has not been demonstrated in prior works, where large pump power is used to assist phase matching and clean OPO pairs typically require sensitive pump detuning. For example, in Ref.~\onlinecite{Fujii2019}, a cluster frequency comb is clearly in competition with the clean OPO pair when the detuning changes. In contrast, our OPO has stable output frequencies. We study this stability by recording the peak amplitude of the 781.5 nm signal versus the pump power dropped inside the microring (Fig. 4(b), inset). The pump threshold power is (1.3 $\pm$ 0.1) mW, similar to, but slightly larger than, that of the previous design. In particular, we only observe one widely-separated OPO pair throughout this detuning process, until at the highest dropped powers, one close-band OPO occurs (Fig.~\ref{Fig4}(c)). Importantly, such close-band OPO, although affecting the output power for the targeted widely-separated OPO, does not come with clustered frequency combs near the signal and idler. We note that an advanced coupling design can help suppress the close-band OPO (see Supplementary Information Section V for details).

The stability can be attributed to three factors. First, our device has a smaller size and thus a FSR of $\approx$ 1 THz, whereas Ref. \onlinecite{Sayson2019} has a larger size and a FSR of $\approx$ 100 GHz to 300 GHz. Second, because of the larger material dispersion at the visible wavelength and the larger geometric dispersion of the nanophotonic devices, the modes around OPO pairs are more dispersive and therefore less prone to clustered frequency combs. More importantly, our devices have smaller operation powers and thus smaller parametric gain bandwidths, which further limits the allowable number of competitive OPO processes. Such superior power efficiency and operation stability comes with a sacrifice of frequency tunability. For example, our results typically show only a few pump modes that can generate widely-separated OPO, as shown in Fig.~\ref{Fig2}(b), while previous works possess $\approx$ 10~$\times$ more pump modes for such operation \cite{Sayson2019}. We note that the frequency tunability can be aided with pump power tuning but not temperature tuning (Supplementary Information Section VI). 

\noindent \textbf{Conclusion}
In summary, we propose and demonstrate, for the first time, visible-telecom OPO using silicon nanophotonics, with a signal-idler spectral separation of $\approx$ 190 THz, and a sub-mW threshold power that is two orders of magnitudes smaller than recently reported infrared OPO~\cite{Sayson2019}. Our demonstration represents a major advance for the on-chip generation of coherent visible light. Compatibility with silicon photonics and its accompanying potential for low-cost, scalable fabrication make our approach particularly promising for integrated photonics applications.



\noindent \textbf{Acknowledgements} This work is supported by the DARPA DODOS and NIST-on-a-chip programs. X.L., G. M., Q.L., and A.S. acknowledge support under the Cooperative Research Agreement between the University of Maryland and NIST-PML, Award no. 70NANB10H193.


\noindent \textbf{Additional Information} Correspondence and requests for materials should be addressed to X.L. and K.S.


\clearpage

\onecolumngrid \bigskip
\setcounter{figure}{0}
\setcounter{equation}{0}
\makeatletter
\renewcommand{\theequation}{S\@arabic\c@equation}
\begin{center} {{\bf \large SUPPLEMENTARY
MATERIAL}}\end{center}

\renewcommand{\thefigure}{S\arabic{figure}}

\section{Theoretical estimate of OPO threshold power}

\begin{center}
\begin{figure}[h]
\begin{center}
\includegraphics[width=0.95\linewidth]{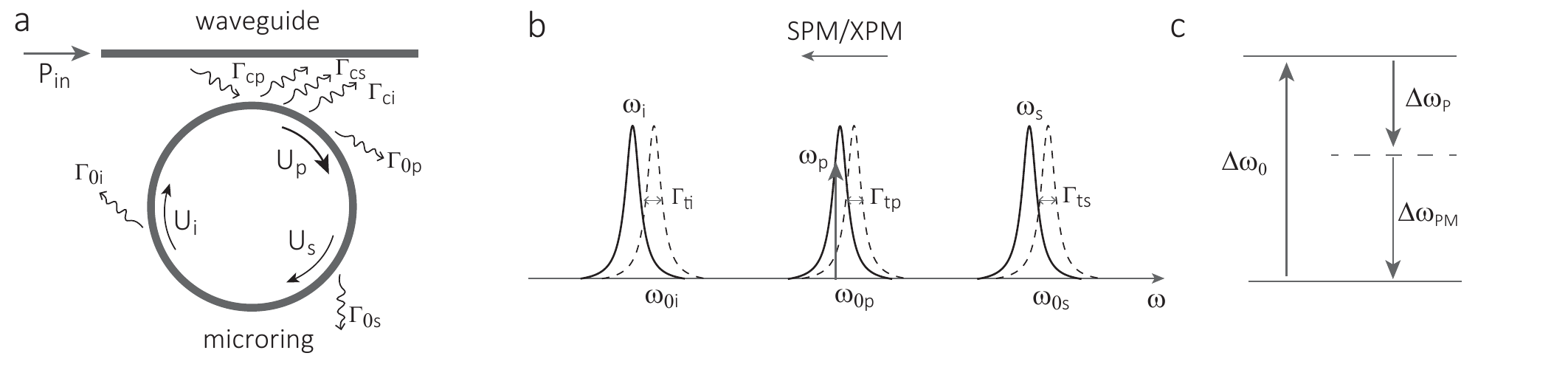}
  \caption{\textbf{Device blueprint and parameters.} \textbf{a,} Nanophotonic OPO scheme. The pump laser with a power of $P_\text{in}$ is coupled at a rate $\Gamma_\text{cp}$ into the microring resonator. The intra-cavity pump energy $U_\text{p}$ is resonantly enhanced by an amount proportional to the photon lifetime in the microring. When the pump energy reaches the threshold value, that is, the four-wave mixing generation rates are larger than the cavity losses for signal and idler modes, intra-cavity OPO ($U_\text{s}$, $U_\text{i}$) can be built up coherently. Each mode has two decay paths in total and the total decay is represented by $\Gamma_\text{tm}$ (m = p,s,i). One decay path is through intrinsic loss of the cavity $\Gamma_\text{0m}$ (e.g., sidewall scattering) and the other path is through coupling out to the waveguide $\Gamma_\text{cm}$. \textbf{b,} Nonlinear resonance shifts of the cavity modes. The intense intra-cavity pump field yields self-phase modulation (SPM) of the pump mode and cross-phase modulation (XPM) of the signal and idler mode, and redshifts the cavity resonances from their natural resonance frequencies $\omega_\text{0m}$ (m = p,s,i). The cavity linewidths remain unchanged and the full-wave-half-maximum (FWHM) values are given by $\Gamma_\text{tm}$. \textbf{c,} Frequency diagram of the OPO process. The OPO process requires pump detuning to compensate the overall frequency mismatch, including natural cavity frequency mismatch and the differences of the SPM and XPM shifts. Here the detunings follow the previous definition $\Delta \omega_\text{m}$ = $\omega_\text{m}$-$\omega_\text{0m}$ (m = p,s,i). The natural frequency mismatch is given by $\Delta\omega_0=\omega_\text{0s}+\omega_\text{0i}-2\omega_\text{0p}$, which is related to $\Delta \nu$ in the main text by a factor of $2 \pi$.}
\label{FigS1}
\end{center}
\end{figure}
\end{center}

In this section, we review optical parametric oscillation (OPO) in high-Q microresonators, and present an estimate of threshold power as a function of the cavity decay rates and effective nonlinearity including mode overlap. In particular, we look into the cases where signal, pump, and idler can be quite different in frequency. In high-Q microresonators, because light propagates many round trips before being lost (e.g., scattering or absorption) or appreciably coupled out from the cavity, we can treat the loss and coupling as if they are uniformly distributed in time and space. The slowly varying light fields satisfy the following equations given in ref.~\onlinecite{Lin2008}:
\begin{eqnarray}
\frac{d\tilde{A}_\text{p}}{dt} = (i \Delta\omega_\text{p} - \Gamma_\text{tp}/2) \tilde{A}_\text{p} + i (\gamma_\text{p} U_\text{p} + 2\gamma_\text{ps} U_\text{s} +2\gamma_\text{pi} U_\text{i}) \tilde{A}_\text{p} + 2i \gamma_\text{pspi} \tilde{A}_\text{s} \tilde{A}_\text{i} \tilde{A}^*_\text{p} + i \sqrt{\Gamma_\text{cp}} \tilde{S}_\text{in}, \label{EqS1} \\
\frac{d\tilde{A}_\text{s}}{dt} = (i \Delta\omega_\text{s} - \Gamma_\text{ts}/2) \tilde{A}_\text{s} + i (\gamma_\text{s} U_\text{s} + 2\gamma_\text{sp} U_\text{p} +2\gamma_\text{si} U_\text{i}) \tilde{A}_\text{s} + i \gamma_\text{spip} \tilde{A}^2_\text{p} \tilde{A}^*_\text{i}, \label{EqS2} \\
\frac{d\tilde{A}_\text{i}}{dt} = (i \Delta\omega_\text{i} - \Gamma_\text{ti}/2) \tilde{A}_\text{i} + i (\gamma_\text{i} U_\text{i} + 2\gamma_\text{ip} U_\text{p} +2\gamma_\text{is} U_\text{s}) \tilde{A}_\text{i} + i \gamma_\text{ipsp} \tilde{A}^2_\text{p} \tilde{A}^*_\text{s},\label{EqS3}
\end{eqnarray}
where $\tilde{A}_\text{m}$ (m = p,s,i) are the intra-cavity light fields for pump, sigal, and idler modes, sitting on the fast-oscillating background of $e^{-i\omega_m t}$, where $\omega_m$ is the angular frequency of the light. Frequency conservation requires $\omega_s+\omega_i=2\omega_p$, which is assumed in deducing the equations. Our convention is to define the higher and lower frequency OPO outputs as signal and idler, respectively. The cavity fields are normalized so that  $|\tilde{A}_\text{m}|^2$ = $U_\text{m}$ (m = p,s,i), which represents the intra-cavity energy. The first terms in Eqs.~(\ref{EqS1}-\ref{EqS3}) describe the free cavity evolution (without sources or nonlinear effects), where $\Delta\omega_\text{m}$ (m = p,s,i) represents the detuning of laser/light frequency ($\omega_\text{m}$) from the natural cavity frequency ($\omega_\text{0m}$), i.e., $\Delta\omega_\text{m}=\omega_\text{m}-\omega_\text{0m}$. $\Gamma_\text{tm}$ describes the decay of the intra-cavity energy $U_\text{m}$, which includes the intrinsic cavity loss and the out-coupling to waveguide, $\Gamma_\text{tm} = \Gamma_\text{0m} +\Gamma_\text{cm}$. Here the decay term $\Gamma_\text{lm}$ is related to optical quality factor $Q_\text{lm}$ by
\begin{eqnarray}
\Gamma_\text{lm} = \frac{\omega_\text{0m}}{Q_\text{lm}},~(l = t,0,c;~m = p,s,i). \label{EqS4}
\end{eqnarray}
We use $\Gamma$ instead of $Q$ so that it is more straightforward to describe the physics of the cavity, as shown in Fig.~\ref{FigS1}(a,b). The second and third terms in Eqs.~(\ref{EqS1}-\ref{EqS3}) describe self/cross-phase modulations (SPM/XPM) and four-wave mixing (FWM) of the cavity fields, respectively. For SPM, $\gamma_\text{m}$ is short for $\gamma_\text{mmmm}$ (m = p,s,i) and describes the phase modulation of the m mode \ks{on} itself. For XPM, $\gamma_\text{mn}$ is short for $\gamma_\text{mnmn}$ (m,n = p,s,i;~ m $\neq$ n) and describes the phase modulation of the m mode by the n mode. The phase modulation, when inside the microring, manifests itself as a shift of cavity frequencies as shown in Fig.~\ref{FigS1}(b). The third-order nonlinear ($\chi^{(3)}$) effects, including SPM, XPM, and FWM, are described by the cavity nonlinear parameter given by the following equation generally:
\begin{eqnarray}
\gamma_\text{mnuv} = \frac{3 \omega_\text{m} \eta_\text{mnuv} \chi^{(3)}_\text{mnuv}}{4 \epsilon_\text{0} \bar{n}^4_\text{mnuv} \bar{V}_\text{mnuv}},~(\text{with~m,n,u,v = p,s,i}), \label{EqS5}
\end{eqnarray}
which is a positive real parameter. $\eta_\text{mnuv}$ characterizes the spatial overlap of interacting optical modes given by:
\begin{eqnarray}
\eta_\text{mnuv} = \frac{\int_\text{V}dv~\sqrt{\epsilon_\text{m}\epsilon_\text{n}\epsilon_\text{u}\epsilon_\text{v}} \tilde{E}^*_\text{m} \tilde{E}_\text{n} \tilde{E}^*_\text{u} \tilde{E}_\text{v}} {(\int_\text{V}dv~\epsilon^2_\text{m} {|\tilde{E}_\text{m}|}^4 \int_\text{V}dv~\epsilon^2_\text{n} {|\tilde{E}_\text{n}|}^4 \int_\text{V}dv~\epsilon^2_\text{u} {|\tilde{E}_\text{u}|}^4 \int_\text{V}dv~\epsilon^2_\text{v} {|\tilde{E}_\text{v}|}^4)^{\frac{1}{4}}}, \label{EqS6}
\end{eqnarray}
where $\tilde{E}_\text{m}$ represents the dominant electric field components of the m = p,s,i mode. This mode is related to $A_\text{m}$ in that $U_\text{m} = |A_\text{m}|^2 \approx \int_\text{V}dv~\epsilon_\text{m} |\tilde{E}_\text{m}|^2$. Here the approximation is made possible when the other electric field componenets are much smaller than the dominant one, for example, $|\tilde{E}_\text{z}|$, $|\tilde{E}_{\phi}|$ $\ll$ $|\tilde{E}_\text{r}|$ for transverse-electric-like (TE) modes.   $\chi^{(3)}_\text{mnuv}$ is short for $\chi^{(3)}(-\omega_\text{m}; \omega_\text{n}, -\omega_\text{u}, \omega_\text{v})$ and represents the third-order nonlinearity at $\omega_\text{m}$ with the inputs at  $\omega_\text{n}$, $\omega_\text{u}$, $\omega_\text{v}$. $\bar{n}_\text{mnuv}$ represents average linear refractive index $\bar{n}_\text{mnuv} = (n_\text{m} n_\text{n} n_\text{u} n_\text{v})^{1/4}$. Likewise, $\bar{V}_\text{mnuv}$ represents average mode volume $\bar{V}_\text{mnuv} = (V_\text{m} V_\text{n} V_\text{u} V_\text{v})^{1/4}$, where individual mode volume is given by:
\begin{eqnarray}
V_\text{m} = \frac{({\int_\text{V} dv~\epsilon_\text{m}|\tilde{E}_\text{m}|}^2)^2}{\int_\text{V}dv~\epsilon^2_\text{m} {|\tilde{E}_\text{m}|}^4}. \label{EqS7}
\end{eqnarray}
The last term in Eq.~(\ref{EqS1}) is the source term that represents the pump laser that is coupled into the cavity. The coupling rate $\Gamma_\text{cp}$ is given by Eq.~\ref{EqS4} and the input field $\tilde{S}_\text{in}$ is normalized in such a way that $|\tilde{S}_\text{in}|^2 = P_\text{in}$ represents the input power in the waveguide (Fig.~\ref{FigS1}).

We note that terms representing phenomena such as nonlinear absorption and free carrier effects are not considered in Eqs.~(\ref{EqS1}-\ref{EqS3}), as silicon nitride (Si$_3$N$_4$) is a wide bandgap material and does not have such effects in the frequency ranges of interest in this work. Moreover, quantum fluctuation of the signal and idler bands are not included because we are only interested in the classical regime, when the OPO is near and above parametric threshold.

We first study the near-threshold case of the OPO process for Eqs.~(\ref{EqS1}-\ref{EqS3}). When the signal and idler intra-cavity energies are small ($U_\text{s}, U_\text{i} \ll U_\text{p}$), their nonlinear contribution can be neglected. The equations are then reduced to
\begin{eqnarray}
\frac{d\tilde{A}_\text{p}}{dt} = {[i(\Delta\omega_\text{p} + \gamma_\text{p} U_\text{p}) - \Gamma_\text{tp}/2]}~\tilde{A}_\text{p}+ i \sqrt{\Gamma_\text{cp}} \tilde{S}_\text{in}, \label{EqS8} \\
\frac{d\tilde{A}_\text{s}}{dt} = {[i(\Delta\omega_\text{s} + 2\gamma_\text{sp} U_\text{p}) - \Gamma_\text{ts}/2]}~\tilde{A}_\text{s} + i \gamma_\text{spip} \tilde{A}^2_\text{p} \tilde{A}^*_\text{i}, \label{EqS9} \\
\frac{d\tilde{A}_\text{i}}{dt} = {[i(\Delta\omega_\text{i} + 2\gamma_\text{ip} U_\text{p}) - \Gamma_\text{ti}/2]}~\tilde{A}_\text{i} + i \gamma_\text{ipsp} \tilde{A}^2_\text{p} \tilde{A}^*_\text{s},\label{EqS10}
\end{eqnarray}
where SPM and XPM terms, now combined with the linear detuning terms, effectively create nonlinear detunings for all three intra-cavity light fields. If we assume that signal and idler fields are nonzero ($|\tilde{A}_\text{s}|, |\tilde{A}_\text{i}| \neq 0$), in steady-state these equations yield
\begin{eqnarray}
{[(\Delta\omega_\text{p} + \gamma_\text{p} U_\text{p})^2 + (\Gamma_\text{tp}/2)^2]}~U_\text{p} = \Gamma_\text{cp} P_\text{in}, \label{EqS11} \\
(\Delta\omega_\text{s} + 2\gamma_\text{sp} U_\text{p})/\Gamma_\text{ts} = (\Delta\omega_\text{i} + 2\gamma_\text{ip} U_\text{p})/\Gamma_\text{ti}, \label{EqS12} \\
(4 \gamma_\text{sp} \gamma_\text{ip}-\gamma_\text{spip} \gamma^*_\text{ipsp} ) U^2_\text{p} + 2(\gamma_\text{sp} \Delta\omega_\text{i}+\gamma_\text{ip} \Delta\omega_\text{s}) U_\text{p} + \Delta\omega_\text{s} \Delta\omega_\text{i} + (\Gamma_\text{ts}/2) (\Gamma_\text{ti}/2)= 0,\label{EqS13}
\end{eqnarray}
where aforementioned energy and power normalizations are used. Here Eq.~(\ref{EqS11}) describes the relation of pump threshold power in the waveguide and the intra-cavity threshold energy. Eq.~(\ref{EqS12}) indicates that the ratios of overall detunings to the cavity linewidths are identical. This equation is related to the coherence of the OPO. Eq.~\ref{EqS13} is the equation leads to the intra-cavity threshold energy, which needs to have a positive solution for OPO to occur. At this point, it is important to recall the requirements for detunings:
\begin{eqnarray}
\omega_\text{s} + \omega_{i} - 2\omega_{p} = \Delta\omega_\text{s} + \Delta\omega_\text{i} - 2\Delta\omega_\text{p} + \omega_\text{0s} + \omega_{0i} - 2\omega_{0p} = 0, \label{EqS14}
\end{eqnarray}

We consider the case that frequency matching can be be perfectly matched for signal and idler, that is, $\Delta\omega_\text{s} = -2 \gamma_\text{sp} U_\text{p}$ and $\Delta\omega_\text{i} = -2 \gamma_\text{ip} U_\text{p}$, which is clearly a solution for Eq.~(\ref{EqS12}). The frequency matching for pump, however, is not guaranteed to be matched automatically and needs to be adjusted by laser detuning. Such detuning needs to consider both the natural frequency mismatch and also the nonlinear cavity shifting due to phase modulation (Fig.~\ref{FigS1}(b)). For simplicity, we define $\Delta\omega_{0} \equiv \omega_\text{0s} + \omega_\text{0i} - 2 \omega_\text{0p}$, which is related to the frequency mismatch ($\Delta \nu$) in the main text by $\Delta\omega_{0} = 2 \pi \Delta \nu$, where when close to pump, positive values correspond to anomalous dispersion and negative values correspond to normal dispersion. We also define $\bar\Gamma_\text{t} \equiv \sqrt{\Gamma_\text{ts}\Gamma_\text{ti}}$, $\gamma_\text{FWM} \equiv \sqrt{\gamma_\text{spip}\gamma^*_\text{ipsp}}$, $\gamma_\text{XPM} \equiv \gamma_\text{sp}+\gamma_\text{ip}$, and $\gamma_\text{PM} \equiv \gamma_\text{sp}+\gamma_\text{ip}-\gamma_\text{p}$. Eqs.~(\ref{EqS13},\ref{EqS14}) then reduces to
\begin{eqnarray}
U_\text{p} = \frac{\bar\Gamma_\text{t}}{2} \frac{1}{\gamma_\text{FWM}}, \label{EqS15} \\
\Delta\omega_\text{p} = \frac{\Delta\omega_{0}}{2} - \frac{\bar\Gamma_\text{t}}{2} \frac{\gamma_\text{XPM}}{\gamma_\text{FWM}}, \label{EqS16}
\end{eqnarray}
We bring these into Eq.~(\ref{EqS11}), and the pump threshold power is therefore given by:
\begin{eqnarray}
P_\text{in} = \frac{1}{\Gamma_\text{cp}} \frac{\bar\Gamma_\text{t}}{2} \frac{1}{\gamma_\text{FWM}} [(\frac{\Delta\omega_{0}}{2} - \frac{\bar\Gamma_\text{t}}{2} \frac{\gamma_\text{PM}}{\gamma_\text{FWM}})^2 + (\frac{\Gamma_\text{tp}}{2})^2], \label{EqS17}
\end{eqnarray}
We can see that the threshold power critically depends on frequency matching. For example, an OPO with an overall detuning of $3 \Gamma_\text{tp}$ would require $\approx$ 10 $\times$ higher threshold power than the ideal case, if all other parameters are the same. In the main text, we search for devices that have zero frequency mismatch ($\Delta\omega_{0} = 0$) for the natural cavity frequencies for convenience. However, this dispersion condition is not optimized for the threshold power. In fact, it is $\Gamma_\text{ti}\Gamma_\text{ts}/\Gamma^2_\text{tp}+1$ times of the ideal case, if we assume the nonlinear parameters are similar for phase matching ($\gamma_\text{sp}$, $\gamma_\text{ip}$, $\gamma_\text{p}$) and four wave mixing ($\gamma_\text{spip}$, $\gamma_\text{ipsp}$). We still use the natural frequency mismatch for two reasons. First, this factor ($\Gamma_\text{ti}\Gamma_\text{ts}/\Gamma^2_\text{tp}+1$) is typically within 2 and therefore does not make a very significant difference. Second, it is difficult in practice to estimate the phase modulation terms accurately \textit{a priori}. Additional simulation data on the dispersion engineering for this natural frequency matching condition are discussed later in Section II and III.

Ideally, the overall detuning should be zero to minimize the threshold power. In another words, the natural frequency mismatch should be a positive value that matches the difference of the SPM and XPM ($\Delta\omega_{0} = \bar\Gamma_\text{t} \gamma_\text{PM}/\gamma_\text{FWM}$). In such case, the threshold power is reduced to
\begin{eqnarray}
P_\text{in} = \frac{1}{\Gamma_\text{cp}} \frac{\bar{\Gamma}_\text{t}}{2} {(\frac{\Gamma_\text{tp}}{2})}^2 \frac{1}{\gamma_\text{FWM}}
= \frac{\omega^2_\text{p} Q_\text{cp}}{Q^2_\text{tp}\sqrt{Q_\text{ts}Q_\text{ti}}} \frac{\epsilon_\text{0}n^2_\text{0p}n_\text{0s}n_\text{0i} \bar{V}_\text{spip}}{6\eta_\text{spsi} \sqrt{\chi_\text{spip}\chi^*_\text{ipsp}}} \label{EqS18}.
\end{eqnarray}
This equation can give some important hints for OPO competition besides frequency matching. First, only pump frequency, but not signal and idler frequencies, is present in this equation, which implies that widely-separated OPOs are no different than close-band OPOs and therefore can be as effective when optimized. Second, this equation suggests that we can suppress the close-band OPO by controlling the coupling Q. We notice that for the close-band OPOs, $Q_\text{ts}$, $Q_\text{ti}$ $\approx$ $Q_\text{tp}$ in general, because both intrinsic and coupling Q values are similar. Therefore, the Q dependence of the threshold power for the close-band OPO is $Q_\text{cp}/Q^3_\text{tp}$, compared to $Q_\text{cp}/(Q^2_\text{tp}\sqrt{Q_\text{ts} Q_\text{ti}})$ for the widely-separated OPO. We define the suppression ratio to be the ratio of these two values, i.e., $\sqrt{Q_\text{ts}Q_\text{ti}}/Q_\text{tp}$. To suppress the close-band OPO relative to widely-separated OPO, we need to increase $Q_\text{ts}$, $Q_\text{ti}$ and decrease $Q_\text{tp}$. Moreover, Eq.~\ref{EqS18} suggests that the threshold power is minimized when the pump is critically coupled ($Q_\text{cp} = Q_\text{0p}$), and signal and idler extremely under coupled ($Q_\text{cs} = Q_\text{ci} = \infty$). In this case, the threshold power is reduced to
\begin{eqnarray}
P_\text{in} = \frac{\omega^2_\text{p}}{Q_\text{0p}\sqrt{Q_\text{0s}Q_\text{0i}}} \frac{2\epsilon_\text{0}n^2_\text{0p}n_\text{0s}n_\text{0i} \bar{V}_\text{spip}}{3\eta_\text{spsi} \sqrt{\chi_\text{spip}\chi^*_\text{ipsp}}} \label{EqS19}.
\end{eqnarray}

\begin{center}
\begin{figure}[h]
\begin{center}
\includegraphics[width=0.8\linewidth]{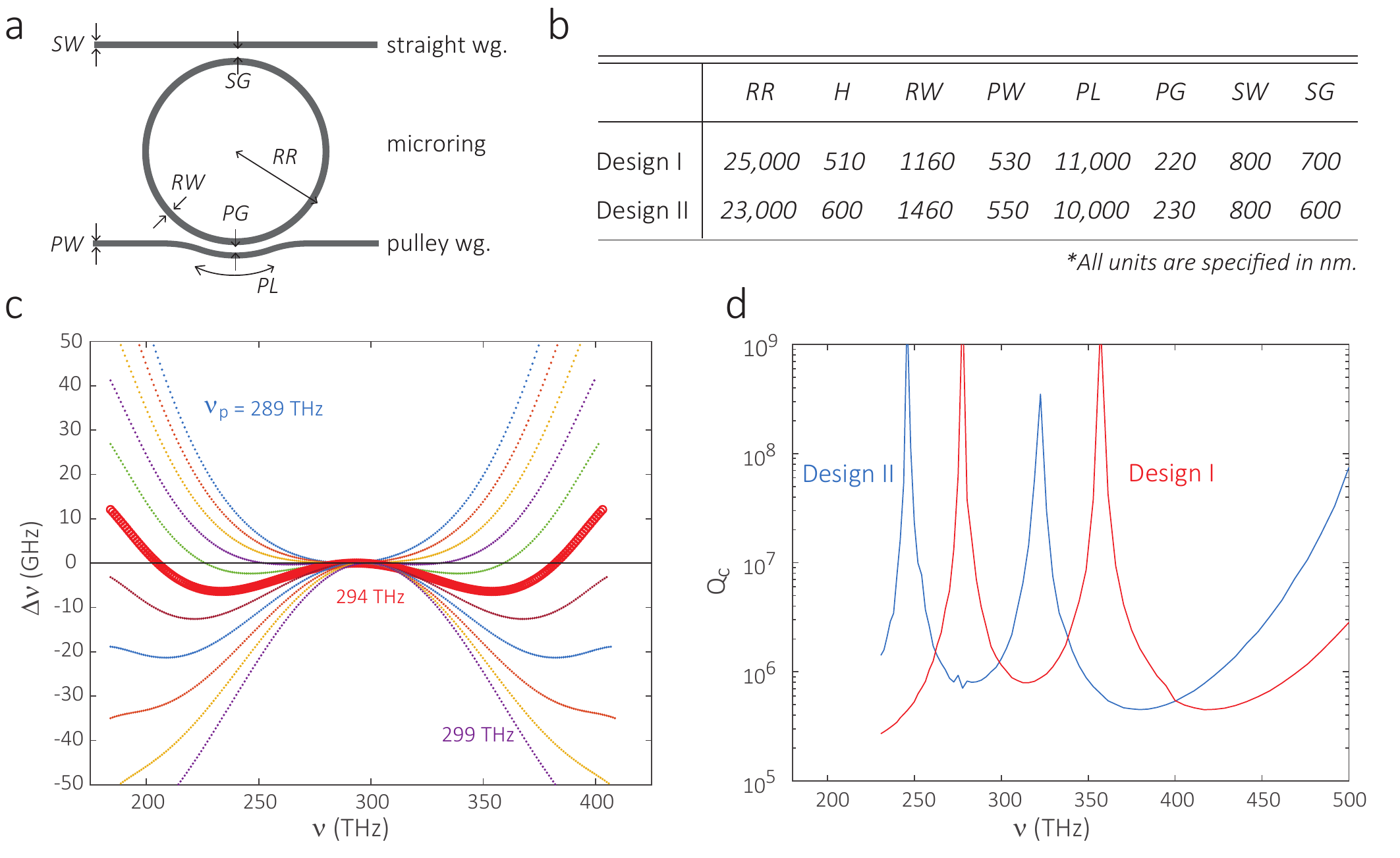}
  \caption{\textbf{Device blueprint and parameters.} \textbf{a,} Device blueprint for widely separated OPO. Two waveguides are used to couple pump, signal, and idler light. The top straight waveguide couples idler light. The bottom pulley waveguide couples pump and signal light. Three parameters control the dispersion of the microring: thickness ($H$), microring width ($RW$), and microring radius ($RR$). Two more parameters are needed for defining the coupling to the straight waveguide(wg.): waveguide width ($SW$) and gap ($SG$). Additionally, three parameters are needed to describe pulley wg. coupling: pulley width ($PW$), pulley gap ($PG$), and pulley length ($PL$). \textbf{b,} A parameter table for two typical geometries studied in the main text. \textbf{c,} The simulated frequency mismatch ($\Delta \nu$) of design II with different pump frequencies. The data of the widely-separated OPO studied in the main text are highlighted in red. \textbf{d,} The simulated frequency-dependent coupling Q ($Q_\text{c}$) of the pulley waveguides in design I and II. We tailor the pulley coupling so that the pulley resonances (dips in coupling Q) fit the pump and visible frequencies for each design.}
\label{FigS2}
\end{center}
\end{figure}
\end{center}

While it is difficult to achieve different coupling for pump, signal and idler modes in the close-band OPOs, it is possible to design such coupling for the widely-separated OPOs, as signal and idler frequencies are separated far way. Moreover, we notice that such configuration of Q factors not only yields the optimized threshold power for the widely-separated OPO, but also naturally suppresses the close-band OPO processes over widely-separted OPO processes. The suppression ratio is $2 \sqrt{Q_\text{0s}Q_\text{0i}}/Q_\text{0p}$, and is $\approx$ 2 assuming $Q_\text{0s}$, $Q_\text{0i}$ $\approx$ $Q_\text{0p}$. The suppression ratio can be further increased when the pump modes are overcoupled, at a price of increased threshold power. Although the coupling effects are generally difficult to isolate to confirm such a suppression ratio in practice, in section IV, we show some experimental data supporting this argument.

In Eq.~(\ref{EqS19}), the parameters to optimize (besides frequency matching and coupling Q engineering) appear in the right term. The refractive indices and mode volumes of the three modes together have a minor difference on the threshold power ($<10\%$) for the widely-separated case and the close-band case (the frequency dependence of $\chi^{(3)}$ is not studied in this paper, and requires further examination). In particular, mode overlap ($\eta_\text{spip}$) is important to guarantee that the widely-separated OPO process be in consideration when competing processes are also potentially realizable. In our case, because all the modes are single fundamental transverse-electric (TE1) modes, the mode overlap is $\> 90\%$ of the perfect case (i.e., close-band OPOs), even when signal and idler are separated $\approx$ 200 THz away.

In summary, analysis of the equations presented in this section shows that widely-separated OPOs can operate at similar threshold powers as close-band OPOs when optimized. We find that, besides dispersion design for frequency matching, coupling quality factor engineering (i.e., through the coupling design) can also be used to optimize the wide-band OPO process, by suppressing the close-band OPOs (see Section IV for details) and/or minimizing the threshold power of the wide-band process.
\vspace{0.2in}

\section{Device parameters: dispersion and coupling}

In the main text, we have used two designs to demonstrate widely-separated OPOs. Here we summarize the devices parameters in Fig.~\ref{FigS2} for both designs. Design I generates OPO at 700 nm and 1300 nm by 920 nm pump. Design II generates OPO at 780 nm and 1500 nm by 1020 nm pump. The device parameters are labeled in Fig.~\ref{FigS2}(a) and their typical values are summarized in Fig.~\ref{FigS2}(b). There are three parameters for the device dispersion - thickness ($H$), ring radius ($RR$), and ring width ($RW$). The dispersion engineering has been already been discussed in detail for design I in the main text. Here we provide further data for design II in Fig.~\ref{FigS2}(c), where the pump is tuned from 289 THz to 299 THz. The device has a radius of 23 $\mu$m and the free spectral range (FSR) is close to 1 THz. When the pump is below 291 THz, we can see that the overall dispersion is anomalous, which is only suited for close-band OPO generation. When the pump frequency is between 292 THz to 294 THz, the dispersion around the pump is normal and the signal and idler are widely separated in frequency. While our simulation range is too small to conclude for the 295 THz case, for pump frequencies above 296 THz, the device seems to be too normal to support any frequency and phase matched modes. The overall trend is similar to design I, but there are only one pair of widely separated modes supported by this design.

In terms of the coupling, we use two waveguides to couple pump, signal, and idler modes, because it is very challenging to couple widely-separated signal and idler within one waveguide. For example, we provide coupling data for design I in Fig.~\ref{FigS2}(d). We use a straight waveguide to couple the idler mode that has the largest wavelength. Because of the evanescent coupling nature, pump and signal modes are more confined within microring and waveguide and therefore are not coupled efficiently by such a waveguide. We also use a pulley waveguide, which is a waveguide with a constant width ($PW$) wrapped around the microring with a constant gap ($PG$) for a certain coupling length ($PL$). Such a structure can efficiently couple pump and signal despite the limited evanescent overlap, because of the increased coupling length, while the waveguide width is chosen so that it is cut off slightly below the idler wavelength and therefore does not couple the idler mode. In Fig.~\ref{FigS2}(d), we show the calculated wavelength-dependent coupling behaviour of our pulley designs. The coupling curves each have two dips (optimal coupling rates) at 325 THz (950 nm) and 450 THz (714 nm) for design I, and 285 THz (1050 nm) and 380 THz (790 nm) for design II, respectively.

\section{Parameter sensitivity for frequency matching}

\begin{center}
\begin{figure}[b!]
\begin{center}
\includegraphics[width=0.85\linewidth]{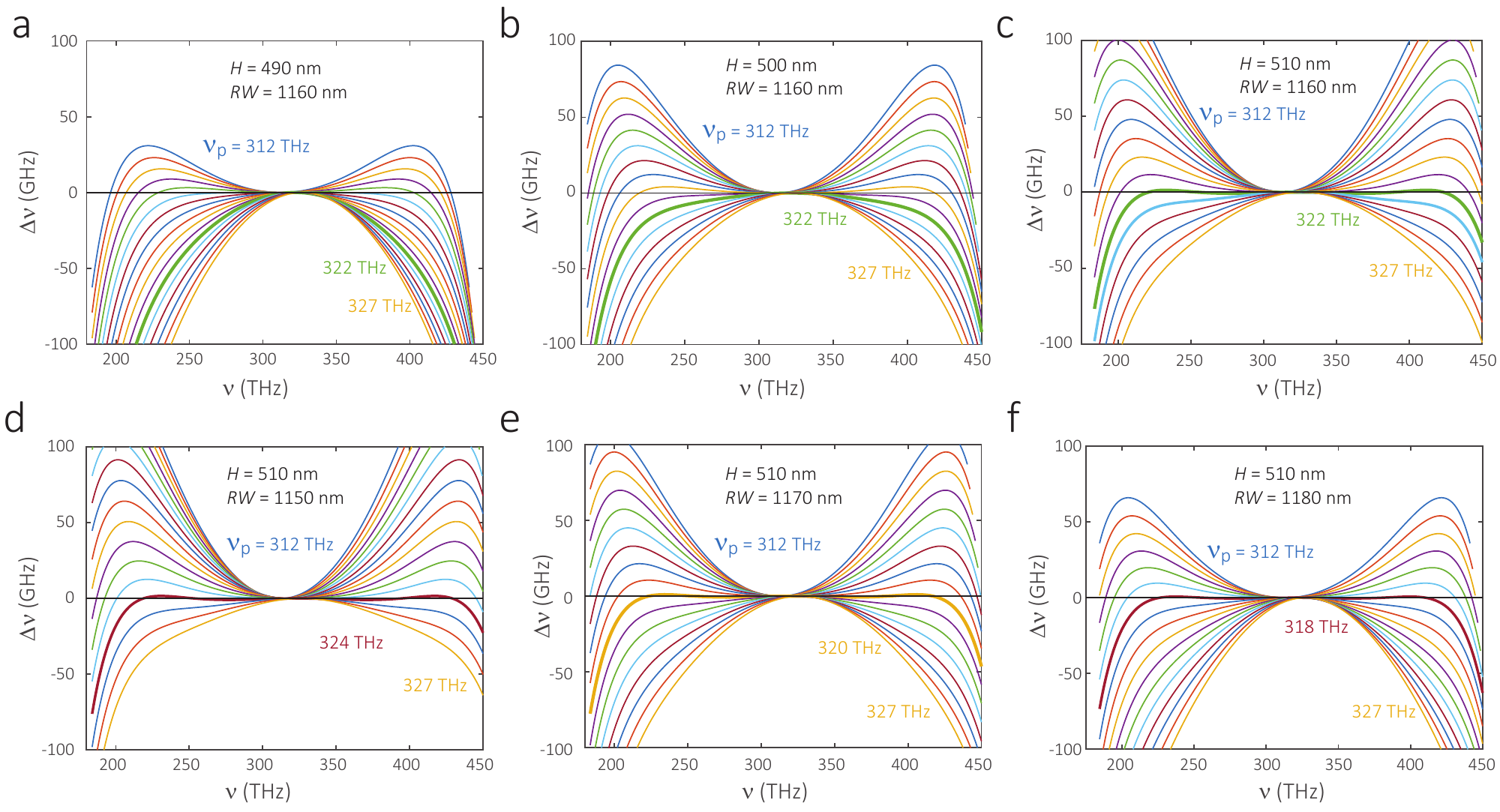}
  \caption{\textbf{Dispersion sensitivity of the frequency match ($\Delta \nu$) on thickness ($H$) and ring width ($RW$).} \textbf{a-c} Dispersion sensitivity on $H$. The widely-separated OPO is very sensitive to $H$. For example, for a fixed $RW$ of 1160 nm, a design with $H$ = 510 nm shows widely-separated phase and frequency matched modes with normal dispersion at the pump (c), while designs with $H$ = 500 nm and $H$ = 490 nm show no widely-separated modes for OPO. $\nu_\text{p}$ changes from 312 THz to 327 THz (from top to bottom) with $\approx$~0.9~THz FSR. The bold green lines indicate $\nu_\text{p} = $  322 THz, which is close to what is investigated in experiment. \textbf{d-f} Dispersion sensitivity on $RW$. The sensitivity of the dispersion to $RW$ is significantly less when compared with $H$. For example, (d,c,e,f) show devices with $RW$ of (1150 nm, 1160 nm, 1170 nm, 1180 nm), respectively. All devices show widely-separated frequency matched modes, with $\nu_\text{p}$ of 324 THz, 322 THz, 320 THz, and 318 THz, respectively. Here all parameters other than $H$ and $RW$ are given in the previous section (design I).}
\label{FigS3}
\end{center}
\end{figure}
\end{center}

In the main text, we discuss two major dispersion design principles for widely-separated OPOs, i.e., phase and frequency matching for the widely-separated mode set and normal dispersion at the pump mode. These two design principles, when separated, have both been achieved previously. It is well known that normal dispersion at one wavelength can be achieved by various parameters, because the change in $H$ can usually be compensated by the change in $RW$. For widely-separated frequency matching only, the dispersion design also shows a similar trend although the design is less trivial \cite{Lu2019B}. In this section, we emphasize that the OPO device is more sensitive in $H$ than $RW$ as the combination of these two principles is nontrivial. For the dispersion engineering based on $H$, we showcase its sensitivity in Fig.~\ref{FigS3}(a)-(c). Here all three devices with $H$ of 490 nm, 500 nm, and 510 nm can satisfy widely-separated frequency matching similarly, but only the 510 nm device (Fig.~\ref{FigS3}(c)) can support the normal dispersion when the pump is at 322 THz (green). In contrast, the first two devices, as shown in Fig.~\ref{FigS3}(a,b), although have normal dispersion at pump frequencies around 322 THz (green), do not support frequency matched pairs for widely-searated OPO for these pump frequencies. Moreover, when these two devices have such widely-separated matched pairs, e.g., when pump is around 315 THz (purple), the dispersion near the pump is quite anomalous. Because of this anomalous dispersion, these widely-separated OPO, although in principle allowed, are usually took over by close-band OPO in practice. We note that this sensitive dependence on $H$ is quite general in design, although we have only show one case here. On the other hand, when we vary the ring width parameter, e.g., (d) 1150 nm, (c) 1160 nm, (e) 1170 nm, and (f) 1180 nm in Fig.~\ref{FigS3}, both widely-separated modes and normal dispersion around the pump are simultaneously obtainable, although the optimized pump frequency shifts slightly as (d) 324 THz, (c) 322 THz, (e) 320 THz, and (f) 318 THz, respectively. We note that the device is sensitive to $RR$ as well, because the bending dispersion also contributes to the overall dispersion design, particularly for the widely-separated case.

\vspace{0.1in}

\section{Device Fabrication}
The device layout was done with the Nanolithography Toolbox, a free software package developed by the NIST Center for Nanoscale Science and Technology~\cite{coimbatore_balram_nanolithography_2016}. The ${\rm Si_3N_4}$ layer is deposited by low-pressure chemical vapor deposition on top of a 3~${\rm \mu}$m thick thermal ${\rm SiO_2}$ layer on a 100~mm diameter Si wafer. The wavelength-dependent refractive index and the thickness of the layers are measured using a spectroscopic ellipsometer, with the data fit to an extended Sellmeier model. The device pattern is created in positive-tone resist by electron-beam lithography. The pattern is then transferred to ${\rm Si_3N_4}$ by reactive ion etching using a ${\rm CF_4/CHF_3}$ chemistry. The device is chemically cleaned to remove deposited polymer and remnant resist, and then annealed at 1100~${\rm ^{\circ} C}$ in an ${\rm N_2}$ environment for 4 hours. An oxide lift-off process is performed so that the microrings have an air cladding on top while the input/output edge-coupler waveguides have ${\rm SiO_2}$ on top to form more symmetric modes for coupling to optical fibers. The facets of the chip are then polished for lensed-fiber coupling. After polishing, the chip is annealed again at 1100~${\rm ^{\circ} C}$ in an ${\rm N_2}$ environment for 4 hours.

\section{Coupling effects on competing OPO processes}

\begin{center}
\begin{figure}[b!]
\begin{center}
\includegraphics[width=0.8\linewidth]{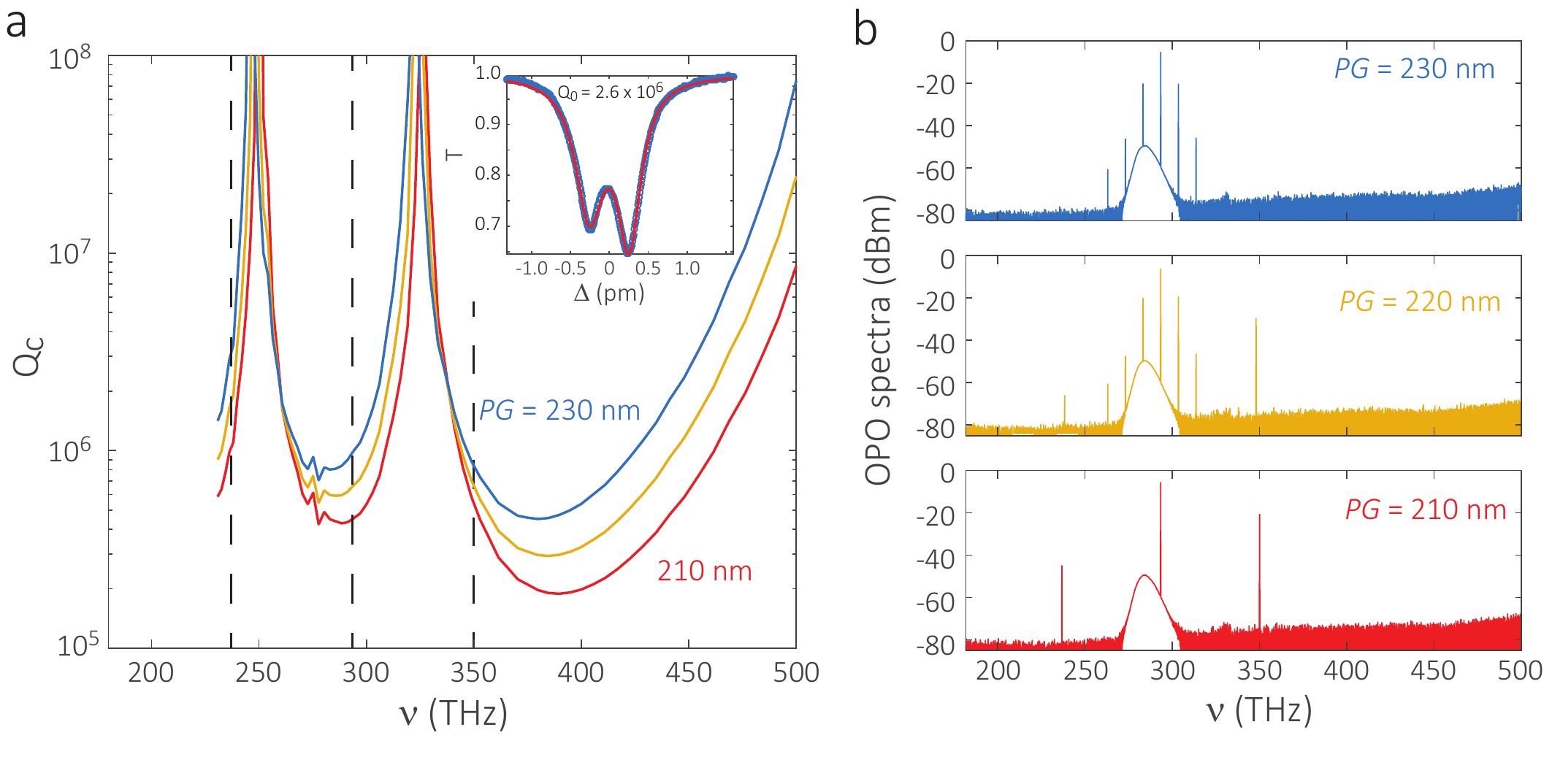}
\caption{\textbf{Waveguide-microring coupling can affect competition between OPO processes.} \textbf{a,} Coupling Qs for pulley gaps of 230 nm, 220 nm, and 210 nm (from top to bottom). The device inset shows a typical optical mode at the pump band. The fitting indicates an intrinsic Q of $(2.6 \pm 0.1) \times 10^6$. Dashed lines indicate the targeted idler, pump, and signal frequencies, from left to right, respectively. \textbf{b,} OPO generation for various pulley gaps. For a closer gap, the close-band OPOs are suppressed because the pump band is more over-coupled in relative to signal and idler band, as expected in Section I.}
\label{FigS4}
\end{center}
\end{figure}
\end{center}

In the main text we proposed that the relative coupling of the pump and signal/idler OPO modes can be used to suppress the close-band OPOs. This idea can also be supported theoretically, see Section I of the Supplementary Information. In this section, we present data in support of this proposal, although the frequencies demonstrated are not as widely separated as those in the optimized devices shown in the main text.

In Section II (Fig.~\ref{FigS2}(d)), we have already shown that the pulley coupling can be optimized to particular pump and signal frequencies. Here we show that when changing the gap of such pulley coupling from 230 nm to 210 nm in 10 nm increments, the change in $Q_\text{c}$ is spectrally non-uniform, as shown in Fig.~\ref{FigS4}(a). We calculate the suppression ratio ($\sqrt{Q_\text{ts}Q_\text{ti}}/Q_\text{tp}$, defined in Secion I) in these cases, where $Q_\text{tm}$ (m = p,s,i) is given by $Q_\text{tm} = Q_\text{0m} + Q_\text{cm}$. $Q_\text{cm}$ is extracted from the simulation (Fig.~\ref{FigS4}(a)) and $Q_\text{0m}$ is assumed to be $2.5~\times~10^6$ (inset of Fig.~\ref{FigS4}(a) shows a fitting of a typical pump transmission recorded experimentally). The suppression ratio is therefore estimated to be 1.31, 1.38 and 1.45 for the gap of 230 nm (blue), 220 nm (yellow), and 210 nm (red). These values suggest that we can have more suppression for the close-band OPO while decreasing the coupling gap, and the trend is clearly observable experimentally in Fig.~\ref{FigS4}(b). In the 230 nm device, only close-band OPOs are observed. In the 220 nm device, both close-band OPO and widely-separated OPO are observed, which indicates that these two OPO cases have similar power thresholds. In the 210 nm device, however, only the widely-separated OPO is observed. We want to emphasize that these devices have the same geometry except for the coupling gap, and are adjacent to each other on the chip so that unintended difference in geometry (e.g., film thickness) are expected to be negligible.

\vspace{0.1in}
\section{Thermal and power effects}
\begin{center}
\begin{figure}[htbp]
\begin{center}
\includegraphics[width=0.95\linewidth]{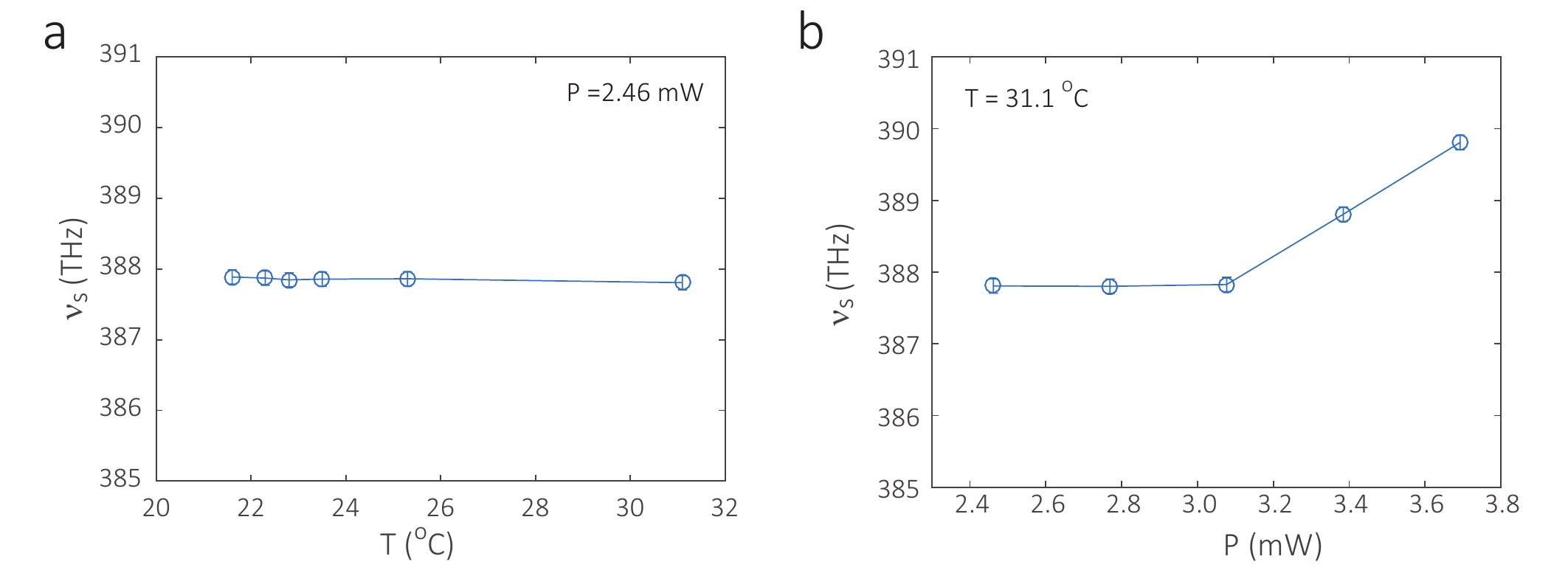}
\caption{\textbf{Thermal and power effects on widely-separated OPOs.} \textbf{a,} Thermal stability of the OPO. At the same pump power level, that is, $P = $~2.46 mW, the device outputs are stable in frequency when temperature (T) changes from 21.5 $^\text{o}$C to 31.1 $^\text{o}$C. \textbf{b,} The OPO outputs are stable for pump power from 2.46 mW to 3.70 mW. When the pump power is further increased, the OPO signal blueshifts 1 FSR for $\approx$~0.3~mW increase. The measurements are carried out at T = 31.1 $^\text{o}$C.}
\label{FigS5}
\end{center}
\end{figure}
\end{center}

In this section, we present experimental data for the thermal dependence and pump power dependence. Figure \ref{FigS5}(a) shows that the OPO frequency is stable over 10 $^\circ$C temperature tuning. This temperature stability allows our device to operate reasonably stable in the environment, also we have not tested the device in extreme temperatures. For the power dependence, we notice that at higher power above the threshold, the device OPO blueshifts to higher frequency as the power increase. While this is not straightforward to estimate numerically, the effect can be explained by Fig.~\ref{FigS1}(c). When the power is so high that the phase change of the intra-cavity energy is larger than the natural cavity mismatch, we require red detuning of the pump, which is usually not directly accessible due to thermal bistability. Therefore, a close mode set with larger natural cavity mismatch becomes the optimized OPO. Here we show that the OPO can be tuned at a rate of 1 FSR (1 THz for 23 $\mu$m device) per 0.3 mW and that the number of FSRs can be adjusted by device radius, in principle.

\end{document}